# A full breakthrough in vacuum ultraviolet nonlinear optical performance of $NH_4B_4O_6F$

Fangfang Zhang[1,2]†, Zilong Chen[1]†, Chen Cui[1]†, Zhihua Yang[1,2], Miriding Mutailipu[1,2], Fuming Li[1], Xueling Hou[1], Xifa Long[1], Shilie Pan[1,2]*

[1]Key Laboratory of Functional Materials and Devices for Special Environmental Conditions, Chinese Academy of Sciences; Xinjiang Key Laboratory of Functional Crystal Materials; Xinjiang Technical Institute of Physics and Chemistry, Chinese Academy of Sciences, Urumqi 830011, China.

[2]Center of Materials Science and Optoelectronics Engineering, University of Chinese Academy of Sciences, Beijing 100049, China.

*Corresponding author. Email: slpan@ms.xjb.ac.cn.

†These authors contributed equally to this work.

**Abstract:** The lack of suitable vacuum ultraviolet (VUV) nonlinear optical (NLO) crystals has hindered the development of compact, high-power VUV sources *via* second harmonic generation (SHG). Here, we report on the development of the fluorooxoborate crystal $NH_4B_4O_6F$ (**ABF**) as a promising material for VUV light generation. For the first time, devices with specific phase-matching angles were constructed, achieving a record 158.9 nm VUV light through phase-matching SHG and a maximum nanosecond pulse energy of 4.8 mJ at 177.3 nm with a conversion efficiency of 5.9 %. The enhanced NLO performance is attributed to optimized arrangements of fluorine-based units creating asymmetric sublattices. This work marks a significant milestone in the field of NLO materials, facilitating the future applications of compact, high-power VUV lasers utilizing **ABF**.

**Main Text:** Compact and efficient vacuum ultraviolet (VUV) light sources in the 100-200 nm range (corresponding to photon energies of 12.4-6.2 eV) are crucial for a wide range of applications, including advanced spectroscopy, quantum research, and semiconductor lithography (*1–3*). Compared to conventional large-scale VUV systems, including synchrotrons, gas discharge, excimer lasers, and free-electron lasers (*4–7*), second-harmonic generation (SHG), discovered in 1961, is recognized as the simplest and most efficient method to produce short-wavelength light *via* frequency-doubling using nonlinear optical (NLO) crystals (*8–10*). However, despite the availability of high-performance NLO crystals (*11–14*), none of these crystals are suitable for VUV SHG applications due to limitations in their material properties. Consequently, sum frequency mixing techniques are employed, which are characterized by their complexity and low efficiency due to the stringent temporal and spatial synchronization required between two laser beams of differing wavelengths (*15, 16*). $KBe_2BO_3F_2$ (KBBF), a benchmark NLO crystal developed in the 1990s, is the only practical material that breaks the "200 nm wall" and enables VUV coherent light generation *via* SHG. Nevertheless, the layered growth habit causes its plate-like morphology along the z-axis, which limits its conversion efficiency, and necessitates a specific prism coupling technique to prevent cutting along the phase-matching direction (*17, 18*). The quest for a suitable VUV NLO material with shorter output wavelengths, higher output energies, and enhanced conversion efficiencies, remains elusive despite ongoing efforts.

Designing VUV NLO materials faces significant challenges, including the combination of conflicting properties such as VUV transparency, strong NLO coefficient, and substantial birefringence for phase-matching at VUV wavelengths. Critically, practical applications require high-quality single crystals of sufficient size for device fabrication, as well as stable physical and chemical properties, high laser-induced damage thresholds (LIDT), and suitable hardness for processing (*19, 20*). Owing to these stringent requirements, no crystal has yet met all these criteria simultaneously. Recently, we proposed a fluorination strategy by substituting fluorine for oxygen atoms in borates to regulate the structure and achieve balanced basic properties that are demanded for VUV applications in a series of fluorooxoborates (*21, 22*). Owing to the limitations of the preparation techniques for large-sized crystals, only preliminary performance assessments have been conducted based on limited millimeter-scale crystals or polycrystalline powders (*23, 24*). In this work, we demonstrated the development and exceptional VUV NLO properties of $NH_4B_4O_6F$ (abbreviated as **ABF**) (*23*) by growing large high-quality single crystals, which reveal superior performances that hold significant promise for practical applications. The fluorination effect on asymmetry and structure motif ordering that leads to performance enhancement are discussed.

**Crystal growth and optical quality of ABF.** The structure of **ABF** features 2D $[B_4O_6F]^\infty$ layers linked by $NH_4^+$ cations through hydrogen bonds with an interlayer distance of 3.81 Å (Fig. 1a). Generally, such a layered structure tends to induce a layered growth habit as observed in KBBF crystals (*17*). We then investigated the interlayer interactions of the structure of **ABF**, which indicates that the insertion of NH$_4^+$ cations with spatial extended s-orbital into the space between neighboring layers leads to relatively large bonding strength, which makes it possible to grow crystals with weak layered tendency (fig. S1). Owing to its growth occurring in a complex, multi-component, gas-liquid-solid three-phase environment, which renders conventional growth methods inefficient, the formation of large-size **ABF** crystals poses a considerable challenge. To address this, we devised an optimized vapor deposition method to produce large single crystals of **ABF**. In contrast to the conventional vapor transport deposition technique that typically requires ultrahigh vacuum conditions and a gas delivery system, our method operates under spontaneous pressure without additional transport agents (See details in Supplementary materials). Fig. 1b shows the as-grown **ABF** crystal with a centimeter scale dimensions of 25 × 18 × 10 mm$^3$. The X-ray rocking curve (Fig. 1c) exhibits a narrow full-width at half-maximum (FWHM) of 36 arcseconds, indicating the high crystalline quality of the sample. The conoscopic interference patterns (Figs. 1d and 1e) unambiguously confirm that **ABF** is biaxial and that the as-grown crystal is optically homogeneous. The thermal expansion coefficients for the $X$, $Y$, and $Z$ axes of **ABF** are measured using $a$-, $c$-, and $b$-oriented crystal samples (Fig. 1f). Like LBO crystals, **ABF** exhibits positive expansion along the $X$ and $Y$ axes and negative expansion along the $Z$ axis. However, its average thermal expansion coefficients ($\alpha_X = \alpha_a = 79.8 \times 10^{-6}\,\text{K}^{-1}$, $\alpha_Y = \alpha_c = 12.1 \times 10^{-6}\,\text{K}^{-1}$, and $\alpha_Z = \alpha_b = -4.16 \times 10^{-6}\,\text{K}^{-1}$) are much smaller than those of LBO ($\alpha_X = 101 \times 10^{-6}\,\text{K}^{-1}$, $\alpha_Y = 31 \times 10^{-6}\,\text{K}^{-1}$, and $\alpha_Z = -71 \times 10^{-6}\,\text{K}^{-1}$) (*26*). The relatively small thermal expansion anisotropy of **ABF** helps prevent cracking during crystal growth and processing. Mechanical hardness measurement on a (100) plate of **ABF** shows a Vickers hardness of 201 (HV0.3, 10 s), corresponding to a moderate Mohs hardness of 4.0, which facilitates processing.

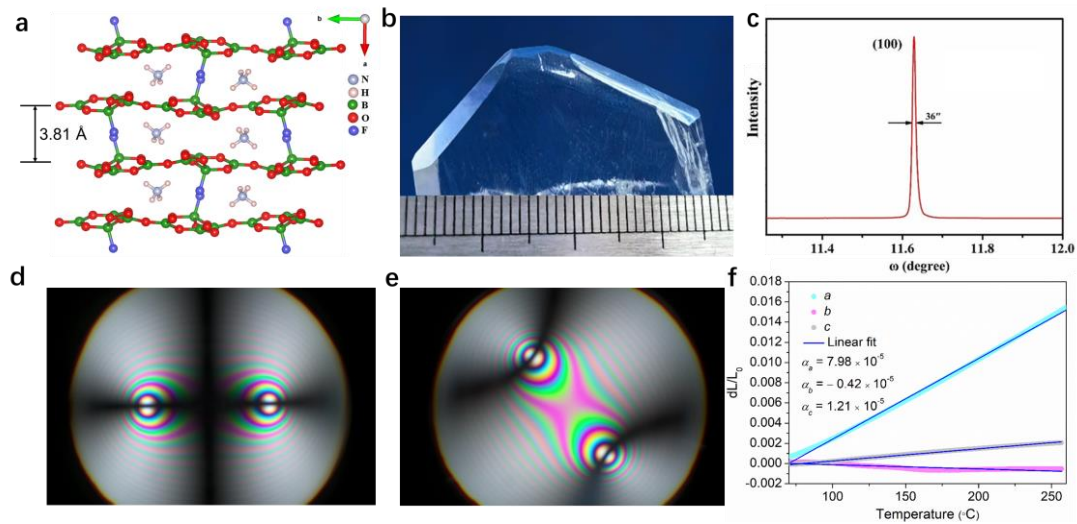

**Fig. 1 | Structure, photograph, optical quality and thermal expansion properties of ABF. a,** Structure of ABF featuring 2D [B$_4$O$_6$F]$^\infty$ layers with an interlayer distance of 3.81 Å. NH$_4^+$ cations are located between the layers forming the final framework *via* hydrogen bonds (not shown for clarity). **b,** As-grown ABF crystal with a centimeter-scale dimensions of 25 × 18 × 10 mm$^3$. **c,** X-ray rocking curve of ABF crystal with a

narrow FWHM of 36″ indicating the high crystalline quality of the sample. **d, e,** Convergent polarized light interference patterns of ABF crystal along the *b* axis (d) and at a 45 ° angle to the *b* axis (e). The black cross and "∞"-shaped interference ring in (d), and the hyperbolic dark bands in (e) indicate the presence of two optic axes in the crystal, thus confirming the biaxial nature of ABF. **f,** Thermal expansion measurements. Based on the well-established relationship $n_X < n_Y < n_Z$, the optical axes (*X, Y, Z*) are determined to correspond to the crystallographic axes (*a, c, b*), respectively. Therefore, the thermal expansion coefficients for the *X, Y,* and *Z* axes of ABF are measured using *a*-, *c*-, and *b*-oriented crystal samples, respectively. It is shown that the thermal expansion ratio of ABF is almost linear over the entire measured temperature range from 70 to 260 °C, and it exhibits positive expansion along the *X* and *Y* axes and negative expansion along the Z axis. The mean linear thermal expansion coefficients in the measured temperature range are calculated according to the thermal expansion ratio curves: $\alpha_X = \alpha_a = 79.8 \times 10^{-6}$ K$^{-1}$, $\alpha_Y = \alpha_c = 12.1 \times 10^{-6}$ K$^{-1}$, and $\alpha_Z = \alpha_b = -4.16 \times 10^{-6}$ K$^{-1}$.

**High performance as a practical VUV NLO crystal**. Proper devices were fabricated from suitable crystals and subjected to performance characterizations that indicate **ABF** is a high-performance crystal for VUV SHG. The main advantages are as follows. First, **ABF** has superior VUV transparency that is crucial for VUV applications. **ABF** exhibits a wide transparency region from VUV to near-IR (Fig 2a). The cut-off edge is down to 155 nm, enabling high transmittance in multiple VUV wavelengths, i.e., 177.3 nm (sixth harmonic generation of 1064 nm Nd: YAG laser) and 193 nm (wavelength of ArF excimer), which are important for both high-resolution photoelectron spectroscopy and photolithography. Second, **ABF** exhibits suitable birefringence and chromatic dispersion to realize phase-matching at VUV wavelengths. The refractive indices are determined by the minimum deviation method and subsequently fit using Sellmeier equations (Fig. 2b). It is confirmed that **ABF** is an optically negative biaxial crystal with $n_z - n_y < n_y - n_x$. In addition, the type I phase-matching curves for SHG are evaluated, indicating that the shortest phase-matching SHG wavelength ($\lambda_{SH}$) for **ABF** is 158 nm in the *XY* plane (Fig. 2c). To the best of our knowledge, **ABF** demonstrates the shortest $\lambda_{SH}$ among the crystals with $\lambda_{SH}$ below 200 nm, as determined by refractive index measurements and dispersion equation fitting (*17, 24, 27–30, 31*). Third, **ABF** exhibits strong SHG effects, essential for high conversion efficiency in the VUV region. **ABF** crystallizes in the space group *Pna*2$_1$ (point group *mm*2), with three non-zero independent second-order NLO coefficients: $d_{31}$, $d_{32}$, and $d_{33}$. For type I SHG (o + o → e) in the *XY* plane, the effective second-order NLO coefficient ($d_{eff}$) can be expressed as $d_{eff} = d_{32}\cos\varphi$, where $\varphi$ is the azimuthal angle (*19*). We then measured $d_{32}$ of **ABF** for SHG from 1064 to 532 nm using the Maker fringe technique, yielding a value of 1.09 pm V$^{-1}$ (Figs. 2d and 2e). Additionally, phase-matching harmonic generation method, conducted under conditions closer to practical applications, reveals a $d_{32}$ value of 0.93 pm V$^{-1}$ (see Supplementary materials). Both tests identify that **ABF** has an extremely large second-order NLO coefficient. The advantage of the SHG effect in **ABF** crystal is more pronounced in the VUV band. By assuming weak wavelength dependence of $d_{ij}$ far from cut-off edge, we deduced $d_{eff}$ across the whole phase-matching range, showing that $d_{eff}$ of **ABF** is consistently much larger than that of KBBF (fig. S2). Specifically, at 386→193 nm and 355→177.3 nm, $d_{eff}$ of **ABF** are 0.63 and 0.48 pm V$^{-1}$, respectively, significantly higher than those of KBBF (0.27 and 0.20 pm V$^{-1}$) at the same wavelengths (*32*). The larger $d_{eff}$ makes **ABF** more favorable for achieving higher laser conversion efficiency, enabling more efficient and powerful laser systems with improved performance. Fourth, **ABF** possesses high LIDT. Fig. 2f shows the LIDT measurement results for **ABF** using LiB$_3$O$_5$ (LBO) (*25*) - the commercial NLO crystal with the highest laser damage resistance - as reference. **ABF** demonstrates a superior LIDT of 1.6 GW cm$^{-2}$ compared to LBO's 1.2 GW cm$^{-2}$ (1064 nm, 8 ns, one-on-one testing), indicating its strong potential for high-power laser applications.

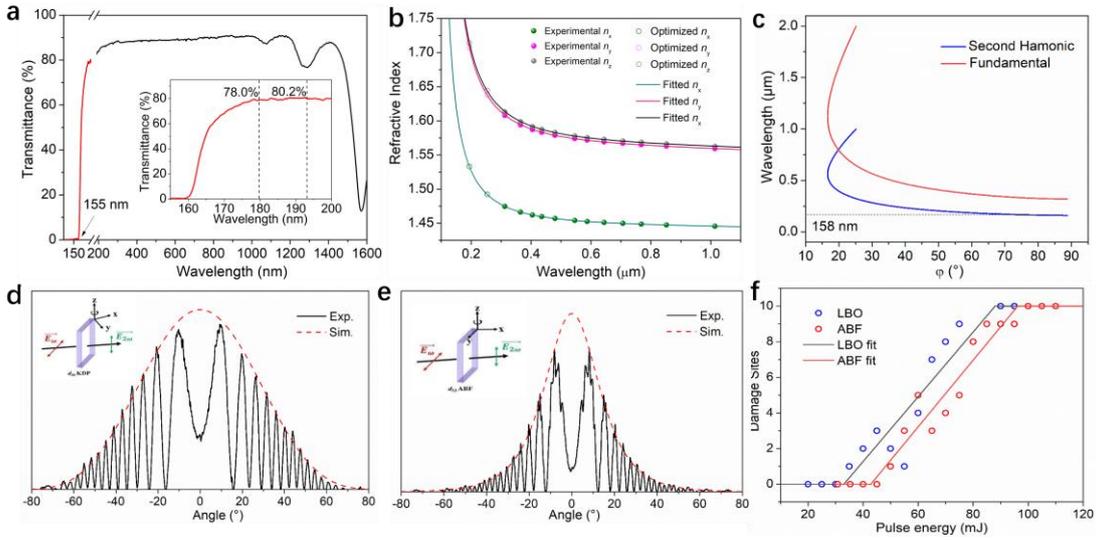

**Fig. 2 | Characterizations of ABF. a,** Transmission spectra of ABF in 155-1600 nm region. Inset: Transmission spectrum in 155-200 nm. ABF exhibits high transmittance (> 85 %) across the broad wavelength range from 200 to 1200 nm. Notable transmittance values include 0.5 % at the cut-off edge of 155 nm, 78.0 % at 177.3 nm, and 80.2 % at 193 nm, all of which are significant for VUV applications. **b,** Refractive index dispersion curves for the ABF crystal. Experimental data are fitted using pentaparametric Sellmeier equations and are further optimized based on the phase-matching angles obtained during the runnable SHG output experiments (see Supplementary materials for details), where $n_x$, $n_y$, and $n_z$ represent the refractive index long *a*-, *c*- and *b*-axes, respectively. It is confirmed that ABF is an optically negative biaxial crystal with $n_z - n_y < n_y - n_x$. **c,** Type I phase-matching curves of the SHG process in the *XY* plane for ABF. **d and e,** SHG measurement using the Maker fringe technique. Measured and calculated Maker fringe data and fitted envelopes for $d_{36}$ of the benchmark crystal KDP (d) and $d_{32}$ of ABF (e) are shown. The data were fitted according to the Maker fringe theory (see Supplementary materials for details). Inset: Measurement configurations for KDP ($\theta = 90°$, $\varphi = 45°$) and ABF ($\theta = 90°$, $\varphi = 0°$). The simulated envelope (red dashed line) indicates that $d_{32}$ of ABF is 2.8 times larger than that of KDP ($d_{36} = 0.39$ pm V$^{-1}$) (*9*), corresponding to 1.09 pm V$^{-1}$. **f,** Comparative

LIDT measurements between ABF and reference LBO crystals. The fitted zero-damage probability fluence corresponds to peak power densities of 1.6 GW cm$^{-2}$ for ABF and 1.2 GW cm$^{-2}$ for LBO (1064 nm, 8ns, one-on one measurement).

**Tunable and high-efficiency frequency doubling outputs.** The harmonic generation in **ABF** crystal is detected using various crystal devices. Tunable frequency-doubled light outputs were achieved from 158.9 to 340.2 nm. Fig. 3 shows the experimental setup and the observation of frequency doubling light for 158.9–188.0 nm (Fig. 3a) using an ABF device with $(\theta, \varphi) = (0°, 70°)$, while other measurements for 190.4–230.5 nm performed on a device with $(\theta, \varphi) = (0°, 40°)$, and for 255.1–340.2 nm performed on a device with $(\theta, \varphi) = (0°, 90°)$ are shown in figs. S4-S7, respectively. The shortest phase-matching SHG output at 158.9 nm surpasses the previous world record of 165 nm held by KBBF crystal (*20*). This breakthrough opens new possibilities in fields such as the study of superconducting mechanisms (where higher photon energies enable a larger observable Brillouin zone, providing critical insights into superconductors) and photochemical investigations (where VUV sources can be employed to probe chemical reactions, as the energy of numerous chemical bonds falls within this wavelength range).

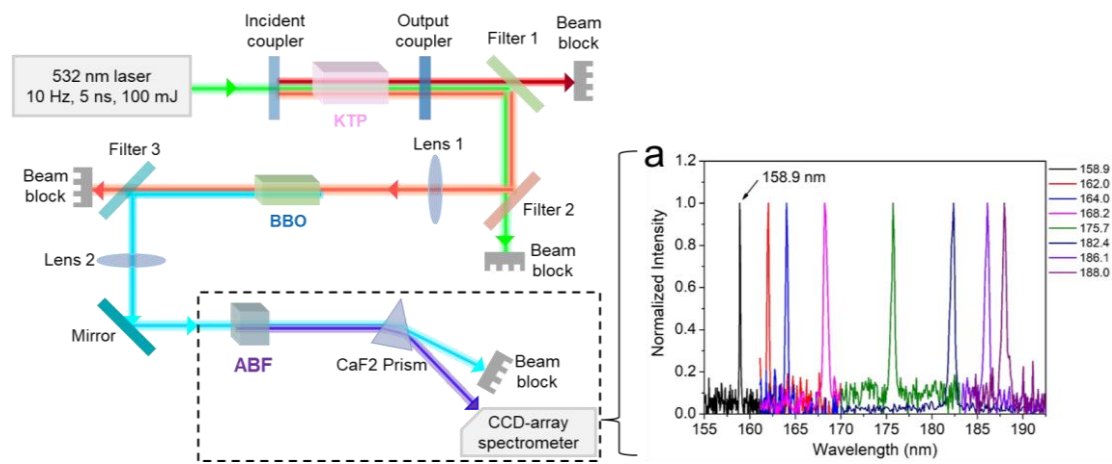

**Fig. 3 | Tunable frequency doubling output of ABF.** Experimental setup for generating tunable nanosecond light at 158.9–340.2 nm is shown. A 532 nm green laser pump (5 ns, 10 Hz, 100 mJ max) drives a KTP-OPO tunable from 620 to 720 nm, and a BBO crystal doubles the KTP-OPO signal to produce the fundamental UV laser for ABF crystal SHG, with UV output below 2 mJ due to bandwidth limitations. Phase-matching is achieved by adjusting the fundamental wavelengths and ABF orientation. All VUV generation processes are conducted in an argon-filled glovebox (indicated by dot lines in the figure). **a,** Observation of frequency-doubling light for 158.9–188.0 nm using an ABF device with $(\theta, \varphi) = (0°, 70°)$. Note that the SHG intensity is normalized for clarity, with raw data presented in fig. S3.

Then, we fabricated SHG devices for the 355→177.3 nm transition to test the frequency doubling ability of **ABF** (Fig. 4). This device circumvents the energy loss inherent in CaF$_2$ or SiO$_2$ prism used in KBBF's prism-coupled devices (Fig. 4a). Output energies at 177.3 nm are recorded for various input energies at 354.7 nm, with results shown in Fig. 4b. An output energy of 4.8 mJ was achieved at a pump energy of 81 mJ, corresponding to an SHG conversion efficiency of 5.9 %, with a maximum efficiency of 7.9 % at 28 mJ. These values significantly exceed those of KBBF (0.375 mJ output energy and 1.76 % conversion efficiency) (*18*). It is worth noting that since the invention of laser technology in 1960, many potential NLO crystals have been reported, but no new crystals have achieved a 177.3 nm SHG output, except for KBBF prism-coupled devices. **ABF** not only achieves this output but also delivers the highest energy output and conversion efficiency to date. Of course, it is anticipated that both the output energy and conversion efficiency could be further improved in future work through enhanced crystal quality and device fabrication precision.

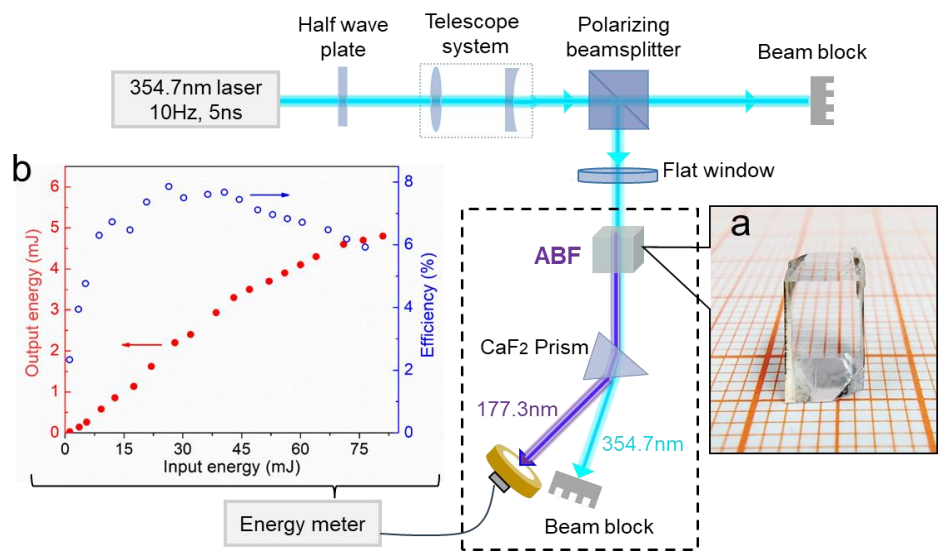

**Fig. 4 High-efficiency frequency doubling output of ABF.** The fundamental laser is a third-harmonic laser emitting at 354.7 nm (5 ns, 10 Hz). Beam optimization is achieved through a telescope system comprising a convex and concave lens, minimizing the beam size before its incidence onto the ABF crystal. The laser beam diameter measures approximately 1.8 mm. VUV laser generation is facilitated by adjusting the ABF crystal's orientation within an argon-filled glove-box. A $CaF_2$ prism, cut at Brewster's angle and positioned behind the ABF crystal, splits the incident laser at 354.7 nm and the generated laser at 177.3 nm. **a,** Device fabricated with phase-matching angle for SHG at 355→177.3 nm. **b,** Output energy and conversion efficiency for SHG at 354.7→177.3 nm.

**Fluorination effect on asymmetry and motif ordering.** NLO performances are primarily determined by the chemistry and electronic structure of local atomic groups. $[BO_3]$ and $[BO_4]$ are two traditional units. However, the $[BO_4]$ unit is rarely favored due to its undesirable contributions to optical response (*33*). The $[BO_3]$ unit plays a dominant role, but non-condensed $[BO_3]$ has dangling bonds which are unfavorable for a large band gap. Thus, few borates containing only $[BO_3]$ anionic groups are VUV NLO materials. In contrast, **ABF** contains both $[BO_3]$ and $[BO_3F]$ groups. The introduction of $[BO_3F]$ unit is beneficial for ABF's full breakthrough in the field of NLO. The electron local function and Laplacian charge density in Fig. 5a show the asymmetrical distribution of [BO₃F] as compared to that of [BO₄] in LBO crystal. In fact, dissimilating tetrahedra can lead to a multi-order character. The Mayer bond order of B-O and B-F bonds in [BO₃F] is approximately 1.28 and 0.63, respectively, illustrating a trend like that of C=C and C-C bonds. The Laplacian charge density between B and O/F further confirms the inequivalent covalent interactions in [BO₃F], which lead to high hyperpolarizability (32.13 a. u.) compared to [BO₄] in LBO (6.93 a. u.). Analysis of the symmetry of **ABF**'s sublattice reveals the absence of symmetry inversion centers for the B and O/F sublattices within the unit cell, indicating non-centrosymmetry (*Pna*2₁, Fig. 5b). Calculations using the symmetry-adapted Wannier functions method (*34, 35*) shows that the contribution to SHG primarily originate from the associated orbitals on the B and O/F sublattices, highlighting the role of noncentrosymmetric sublattice band-edge orbitals. Specifically, the B and O/F orbitals contribute over 77 % to SHG (table S1).

Usually, the phonon (lattice vibration) frequency is in the THz range, which is much lower than the photon energy of incident light (in the eV range). Therefore, we neglect the ionic contribution to the NLO response and consider only the electron-photon interaction with dipole approximation. The total model Hamiltonian is given by,

$$H = h_0 + h_1 = h_0 + e\vec{E}(t) \cdot \vec{r} \quad (1)$$

where $e = 1.6 \times 10^{-19}$ C, and the light field $\vec{E}(t) = E(\omega)e^{-i\omega t} + c.c.$ ($\omega$ is positive, and $c.c.$ denotes complex conjugate). Here, $h_0$ is the single-particle Hamiltonian, and $h_1$ represents the dipole interaction. For simplification, we assume a two-band model with one conduction band $|c\rangle$ with eigenvalue $\varepsilon_c$ and one valence band $|v\rangle$ with eigenvalue $\varepsilon_v$. Band structure calculations for **ABF** indicate that the valence band is dominated by O-2p or F-2p orbitals, while the conduction band is dominated by B-2p orbitals. And the charge polarization $P = -e\langle\psi|r|\psi\rangle$ reads,

$$P(t) = -e\langle v|r|v\rangle + \left[\frac{\langle c|e\vec{E}(t) \cdot \vec{r}|v\rangle}{\hbar\omega - \varepsilon_{cv}}\langle v|r|c\rangle + c.c\right] + P^{(2)}(t) \quad (2)$$

where $\varepsilon_{cv} = \varepsilon_c - \varepsilon_v$ is approximately the band gap (since the band dispersion is minimal, *i.e.*, flat bands). The first term means spontaneous charge polarization, and the second term describes the linear response, and the third term accounts for the nonlinear response. The second-order polarization $P^{(2)}(t)$ within the shift vector mechanism is,

$$P^{(2)}(t) = e \int a_c^\dagger a_c (r_{vv} - r_{cc}) d\vec{k} \quad (3)$$

where shift vector $r_{vv} - r_{cc} = \langle v|r|v\rangle - \langle c|r|c\rangle$ characterizes the difference in position centers between the valence and conduction bands. That is why $P^{(2)}(t)$ in equation (3) is referred to as the shift vector. For crystals lacking spatial inversion symmetry, the shift vector is nonzero. For ionic bonds, $r_{vv} - r_{cc}$ is approximately the bond length. The second-order response $P^{(2)}(t)$ includes both static part and SHG. In **ABF**, the charge transfer between F-B and O-B induces both static and frequency doubling second-order charge polarization.

Fluorine as the bond terminator can "cut" the B–O framework as a scissor, directly regulating the anionic framework, and ensuring a two-dimensional anionic framework with high flatness of the center atoms B (0.75 Å, Fig. 5c). The [BO$_3$F] unit regulates the motif ordering to achieve the high polymerization. The high orientation arrangement of polymerized anionic groups in the layered structure of ABF ensures the effective superposition of polarization anisotropy of [BO$_3$] and [BO$_3$F] (Fig. 5d), which enhances birefringence and phase-matching ability. High polymerization and orientation drive the elimination of the dangling bond of oxygen, facilitating VUV transmission, large birefringence and enhanced NLO response.

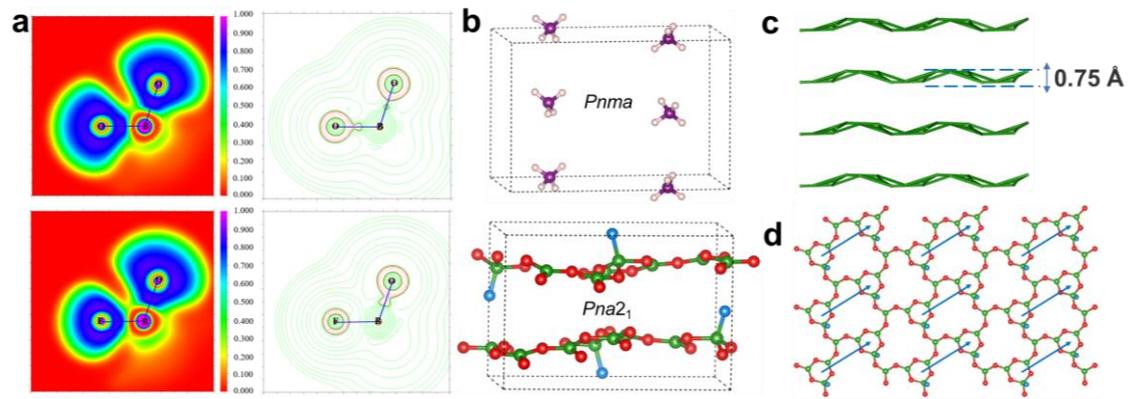

**Fig. 5 | Theoretical analysis. a,** Electron local function (ELF, left) and Laplacian charge density ($\nabla^2\rho$, right) of [BO$_4$] unit for LBO (top) and [BO$_3$F] unit for ABF (bottom). **b,** Symmetry of the sublattice for ABF, where the sublattice of NH$_4^+$ (top) belongs to the centrosymmetric *Pnma* space group and the sublattice of B-O/F (bottom) belongs to the non-centrosymmetric *Pna*2$_1$ space group. **c,** Two-dimensional anionic framework with high flatness of the center B atoms in ABF. **d,** Arrangement of polymerized anionic groups in the layered structure of ABF, and the blue lines represent the high orientation arrangement of microscopic groups.

**Conclusion: ABF** is the first VUV NLO crystal, to our best knowledge, that can generate high-power VUV light through SHG without using a prism coupling technique. Centimeter-sized, high-quality single crystals of **ABF** can be grown using the optimized seeded vapor deposition method, enabling the fabrication of devices with specific phase-matching angles for VUV light. **ABF** exhibits excellent comprehensive performance, including short $\lambda_{\text{cut-off}}$ (155 nm), broad phase-matching range (~ 158 nm), large effective SHG coefficient (0.48 pm V$^{-1}$ at 355 nm), and an extremely high laser damage threshold (1.6 GW·cm$^{-2}$, 1064 nm, 5 ns, 1 Hz). Using an **ABF** device, the shortest VUV wavelength of 158.9 nm, ever generated through SHG in a phase-matching bulk crystal, was detected. An astonishing maximum pulse energy of 4.8 mJ at 177.3 nm was achieved with a high second harmonic conversion efficiency of 5.9 %. The introduction of fluorine to form asymmetric [BO$_3$F] with inequivalent covalent interactions drives the motif ordering and realizes fully upgraded NLO performances.

The full breakthrough of **ABF** paves the way for compact, efficient all-solid-state VUV lasers, promoting their widespread applications in contemporary science and industry.

# Supplementary Materials for

# A full breakthrough in vacuum ultraviolet nonlinear optical performance of $NH_4B_4O_6F$

Fangfang Zhang, Zilong Chen, Chen Cui, Zhihua Yang, Miriding Mutailipu, Fuming Li, Xueling Hou, Xifa Long, Shilie Pan

Corresponding author: slpan@ms.xjb.ac.cn

**fig. S1.** Schematic diagram of the enhancement of interlayer coupling through the insertion of atoms with spatially extended s-orbitals between neighboring layers.

**fig. S2.** Comparison of the effective NLO coefficients ($d_{eff}$) between ABF and KBBF at fundamental wavelength range of 310-400 nm.

**fig. S3.** Raw data of frequency doubling light for 158.9-188.0 nm using the ABF device with ($\theta$, $\varphi$) = (0 °, 70 °).

**fig. S4.** Observation of frequency doubling light for 190.4-230.5 nm using the ABF device with ($\theta$, $\varphi$) = (0 °, 40 °).

**fig. S5.** Raw data of frequency doubling light for 190.4-230.5 nm using the ABF device with ($\theta$, $\varphi$) = (0 °, 40 °).

**fig. S6.** Observation of frequency doubling light from 255.1-340.2 nm using the ABF crystal with ($\theta$, $\varphi$) = (0 °, 90 °).

**fig. S5.** Observation of frequency doubling light for 158.9-188.0 nm using the ABF crystal with ($\theta$, $\varphi$) = (0 °, 70 °).

**fig. S6.** Observation of frequency doubling light for 190.4-230.5 nm using the ABF crystal with ($\theta$, $\varphi$) = (0 °, 40 °).

**fig. S7.** Raw data of frequency doubling light for 255.1-340.2 nm using the ABF device with ($\theta$, $\varphi$) = (0 °, 90 °).

**table S1.** Calculated effective SHG coefficients tensor (pm V$^{-1}$) and contribution of orbitals in the SHG coefficient $d_{32}$ of ABF.

## Materials and Methods

### Characterization:

**Transmittance spectra** were measured on (100) plates without coatings. Data in the wavelength from 200 to 1600 nm were detected on a 2.3 mm-thick plate by SolidSpec-3700DUV spectrophotometer in a nitrogen gas atmosphere. Data in the wavelength from 155 to 200 nm were detected on a 0.05 mm-thick plate by Metrolux ML6500 VUV spectrophotometer under vacuum conditions.

**Laser-induced damage threshold** (LIDT) measurement was carried out on a (100) plate of ABF by using a pulsed laser that has a Gaussian-shaped pulse duration of 10 ns (FWHM at 1064 nm). The UV-SHG (700 → 350 nm) phase-matching LBO crystal was measured by a one-on-one test for the reference. The damage probability was plotted as a function of the fluence, and the data were linearly extrapolated to determine the fluence at which the damage probability is zero, yielding the LIDT value.

**Thermal expansion coefficients** for the X, Y, and Z axes of **ABF** were measured using a-, c-, and b-oriented crystal samples with lengths being 4.42, 4.62 and 3.47 mm, respectively, by a NETZSCH DIL 402 PC dilatometer in the temperature range of 70-260 °C with a heating rate of 5 °C min$^{-1}$ in air.

**Mechanical hardness measurement:** The Vickers hardness of ABF was measured on a (100) plate using a DHV-1000 microhardness meter, with HV0.3 and a dwell time of 10 s. Five points were tested, and the average value was calculated as the final value. Mohs hardness (HM) was calculated from Vickers hardness (HV) by using the following equation: HM = 0.675(HV)1/3.

**Refractive indices** were measured using the minimum deviation method. Two prisms of ABF with apex angles of ~20 ° and different cutting orientations were used. The measurements were performed on 12 different monochromatic sources across a broad wavelength range from 253 to 1013 nm, and subsequently fitted by Sellmeier equations $n_i^2 = A + \dfrac{B}{\lambda^2 - C} - D \times \lambda^2$, where $\lambda$ is the wavelength (unit of $\lambda$ is in μm) and A−D are the parameters. In addition, the Sellmeier equations were further optimized based on the phase-matching angles obtained during the runnable SHG output experiments in the range of 158.9 to 235 nm. This optimization provides a relatively accurate refractive index dispersion equation for ABF over a wide range extending from the VUV to near-infrared spectral region.

**The nonlinear coefficient $d_{32}$ of ABF** was determined using two methods. For the Maker fringe technique, a KDP crystal ($\theta$ = 90 °, $\varphi$ = 45 °) and an ABF crystal ($\theta$ = 90 °, $\varphi$ = 0 °) were employed for measurements. The KDP crystal was rotated on the *XY* plane with *Z* as the axis, while the ABF crystal was rotated on the *XZ* plane with *Y* as the axis. The Maker fringes of both KDP and ABF crystals were obtained.

Phase-matching measurements were conducted employing a high-power pulsed laser operating at 1064 nm, featuring a pulse duration of approximately 5 ns, a repetition rate of 40 kHz, and a maximum power output of 40 W. The laser beam was focused to a diameter of ~ 400 μm at the waist utilizing a lens with a focal length of 750 mm. A LBO crystal ($\theta$ = 90 °, $\varphi$ = 11.3 °) with anti-reflective coating at 1064 nm and 532 nm, and an ABF crystal ($\theta$ = 90 °, $\varphi$ = 0 °) with no coating, were employed to measure the output power of SHG from 1064 to 532 nm at phase-matching orientations. These crystals were successively positioned at the waist of the 1064 nm laser beam. At a pump power of 20 W, the output powers of ABF and the referenced LBO crystals were 127 and 120 mW, respectively, which reveals a value of 0.93 pm V$^{-1}$ for $d_{32}$ of ABF.

**Tunable and high-efficiency frequency doubling output of ABF：**

**Tunable nanosecond harmonic light generation**

The experimental settings for generation of tunable 158.9-340.2 nm light is shown in Fig. 3 in the main text, which comprises a 532 nm green laser pump with a pulse duration of 5 ns and a repetition rate of 10 Hz, capable of delivering a maximum pulse energy of 100 mJ. The optical parametric oscillator (OPO) signal, tunable from 620 to 720 nm via a KTP crystal. The frequency doubling of the KTP-OPO signal by a BBO crystal serves as the fundamental laser for the second-harmonic generation (SHG) of the ABF crystal. Owing to the limited bandwidth of the KTP-OPO signal, the energy output of the tunable UV laser remains below 2 mJ. Collimated by L2 and focused by L3, the UV laser interacts with an ABF crystal positioned at its focus. By adjusting the ABF orientation and the wavelengths of the KTP-OPO signals and BBO-SHG to satisfy phase-matching conditions, the ABF-SHG output is detected with a CCD-array spectrometer. A CaF$_2$ prism, angled at Brewster's angle, splits the fundamental UV beam and the VUV signal. All operations for generating the VUV signal are conducted within an argon-filled glove-box environment.

**Generation of high energy 177.3 nm nanosecond laser in ABF**

The experimental configuration is depicted in Fig. 4 in the main text. The fundamental laser is a third-harmonic laser emitting at 354.7 nm with a pulse duration of 5 ns and a repetition rate of 10 Hz. Beam optimization is achieved through a telescope system comprising a convex and concave lens, minimizing the beam size before its incidence onto the ABF crystal. The laser beam diameter measures approximately 1.8 mm. Within an argon-filled glovebox, VUV laser generation is facilitated by adjusting the ABF crystal's orientation. A CaF$_2$ prism, cut at Brewster's angle and positioned behind the ABF crystal, splits the incident laser at 354.7 nm and the generated laser at 177.3 nm.

**First principles calculations**

The B3LYP (Becke, three-parameter, Lee-Yang-Parr) exchange-correlation functional with the Lee-Yang-Parr correlation functional at the 6-311g basis set in Gaussian (*36*) was employed to calculate the properties of the [BO$_3$F] unit in ABF. And the electronic wave function analysis is based on multiwfn software (*37*). In this paper, we used the Vienna ab initio simulation package (VASP) (*38, 39*) to simulate the first principles, and Perdew-Burke-Enzerhof (PBE) functional and projector-augmented wave (PAW) methods (*40, 41*) to describe the exchange correlation potential and ion-electron interaction, respectively. The outmost electrons of H-1s, B-2s2p, O-2s2p, and F-2s2p are regarded as valence electrons. The Broyden-Fletcher-Goldfarb-Shannon (BFGS) algorithm was used to optimize the atomic position and lattice parameters, and the atoms were allowed to relax until the force applied to the atom is less than 0.02 eV/Å. An energy cutoff 400 eV and a k-point spacing with 0.03 Å$^{-1}$ were used. During the static self-consistent-field calculation, the plane-wave cutoff energy of 600 eV, threshold of 10$^{-7}$ eV and dense Monkhorst–Pack k-point mesh spanning less than 0.03 Å$^{-1}$ were performed. And the Wannier functionals were constructed through a post processing procedure using the output of VASP calculation, and the corresponding orbitals type by the projection of all valence states in the unit cell were generated using WANNIER90 (*42*). The optical properties of all compounds and the SHG contribution of each Wannier orbital were calculated. Here, the plane-wave cutoff energy of 800 eV, threshold of 10$^{-10}$ eV and the dense Monkhorst-Pack k-point grid is twice as static self-consistent-field calculation.

The length gauge formalism method was developed by Aversa and Sipe (*43*) to avoid unphysical divergences. At a zero frequency, the formula of second-order NLO coefficients can be derived as

$$\chi_{ijk}^{(2)} = \chi_{ijk}^{(2)}(\text{VE}) + \chi_{ijk}^{(2)}(\text{VH}) + \chi_{ijk}^{(2)}(\text{TB})$$

Where,

$$\chi_{ijk}^{(2)}(\text{VE}) = \frac{e^3}{2\hbar^2 m^3} \sum_{vcc'} \int \frac{d^3k}{4\pi^3} \, P(ijk) \, \text{Im}[p_{vc}^\alpha p_{cc'}^\beta p_{c'v}^\gamma] \left( \frac{1}{\omega_{cv}^3 \omega_{vc'}^2} + \frac{2}{\omega_{vc}^4 \omega_{c'v}} \right)$$

$$\chi_{ijk}^{(2)}(\text{VH}) = \frac{e^3}{2\hbar^2 m^3} \sum_{vv'c} \int \frac{d^3k}{4\pi^3} \, P(ijk) \, \text{Im}[p_{vv'}^\alpha p_{v'c}^\beta p_{cv}^\gamma] \left( \frac{1}{\omega_{cv}^3 \omega_{v'c}^2} + \frac{2}{\omega_{vc}^4 \omega_{cv'}} \right)$$

Here, *i, j, k* are Cartesian components, v and v′ denote valence bands, c and c′ denote conduction bands, and P (*ijk*) denotes full permutation and explicitly shows the Kleinman symmetry. In addition, the contribution of the two-band (TB) transition process has been strictly proved to be zero (*44*). Using unitary transformation, a set of WFs $w_{n\mathbf{R}}(\mathbf{r}) = w_n(\mathbf{r}-\mathbf{R})$ labeled by Bravais lattice vector **R** can be constructed by using Bloch eigenstates $\psi_{n\mathbf{k}}$ of band *n* (*45*).

$$|R_\alpha\rangle = \frac{V}{(2\pi)^2} \int_{BZ} e^{-i\mathbf{k}\cdot\mathbf{R}} |w_{\alpha\mathbf{k}}\rangle d\mathbf{k}^3$$

For the projection coefficients $C_{n\mathbf{k}}^\alpha$:

$$|n\mathbf{k}\rangle = \sum_\alpha C_{n\mathbf{k}}^\alpha |w_{\alpha\mathbf{k}}\rangle$$

Here, $|C_{n\mathbf{k}}^\alpha|^2$ is the weight of $w_\alpha$ to the *n*th valence band-decomposed SHG $\chi_{ijk,n\mathbf{k}}^{(2)}$.

By rigorous derivation, the contribution of the total SHG $\chi_{ijk,n\mathbf{k}}^{(2)}$ can be written:

$$\chi_{ijk,w_\alpha}^{(2)} = \sum_{|n\mathbf{k}\rangle \in \{v\}} |C_{n\mathbf{k}}^\alpha|^2 \chi_{ijk,n\mathbf{k}}^{(2)}$$

**Interlayer coupling calculation** As shown in fig. S1, if for two atomic orbitals with relatively long distance, saying two $p_z$ orbitals along the z-direction, the hopping parameter,
$$t = \langle p_z|h_0|p_z\rangle = \int \psi_z^*(\vec{r}-\vec{d})h_0(\vec{r})\psi_z(\vec{r}+\vec{d})dr$$
will be minute, and the interlay coupling is dominating by the Van der Waals force. To enhance the interlayer coupling, we should insert some atoms into the space between neighboring layers to form atomic bonding. For example, we insert one s-orbital into the separation, and the hopping parameter between neighboring layers read,
$$t' = (\langle p_z| + \alpha^*\langle s|)h_0(|p_z\rangle + \alpha|s\rangle)$$
$$= \int \psi_z^*(\vec{r}-\vec{d})h_0(\vec{r})\psi_z(\vec{r}+\vec{d})dr + \alpha\langle p_z|h_0|s\rangle + \alpha^*\langle s|h_0|p_z\rangle$$
$$t' = t + \alpha\langle p_z|h_0|s\rangle + \alpha^*\langle s|h_0|p_z\rangle$$
$$\alpha = \frac{\langle s|h_0|p_z\rangle}{\Delta} = \frac{(\mathrm{sp}\sigma)}{\Delta}$$
, where $\Delta$ is the charge transfer energy between $p_z$-orbital and $s$-orbital which characterizes the difference of electronegativity between two different atoms.
$$t' = t + 2\frac{(\mathrm{sp}\sigma)^2}{\Delta}$$
This interlayer hopping parameter indicates that we can insert atom or atomic cluster with spatial extended s-orbital to obtain relatively large bonding strength (spσ).

**fig. S1. Schematic diagram of the enhancement of interlayer coupling through the insertion of atoms with spatially extended s-orbitals between neighboring layers.** It illustrates the modified hopping parameter and the resulting increase in bonding strength, which enables the growth of crystals with reduced layered tendency.

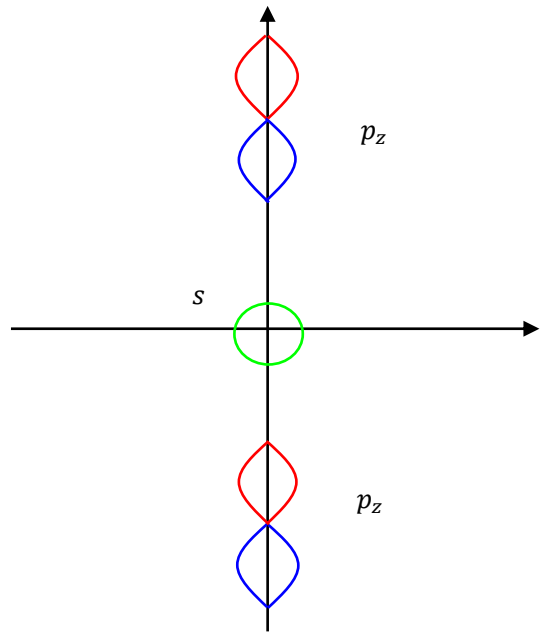

**fig. S2. Comparison of the effective NLO coefficients ($d_{\text{eff}}$) between ABF and KBBF at fundamental wavelength range of 310-400 nm.** For type I phase-matching, $d_{\text{eff}}$ (ABF) = $d_{32}\cos\varphi$, where $d_{32}$ = 1.09 pm V$^{-1}$ based on the Maker fringe technique, and $d_{\text{eff}}$ (KBBF) = $d_{11}\cos\theta\cos3\varphi$, where $d_{11}$ = 0.49 pm V$^{-1}$ (*32*). The phase-matching angle ($\theta$) and azimuthal angle ($\varphi$) were calculated based on the Sellmeier equations for ABF and KBBF, respectively. The weak dependence of $d_{ij}$ on wavelength where far from cut-off edge was ignored. $d_{\text{eff}}$ at 386→193 nm and 355→177.3 nm is highlighted by dot lines.

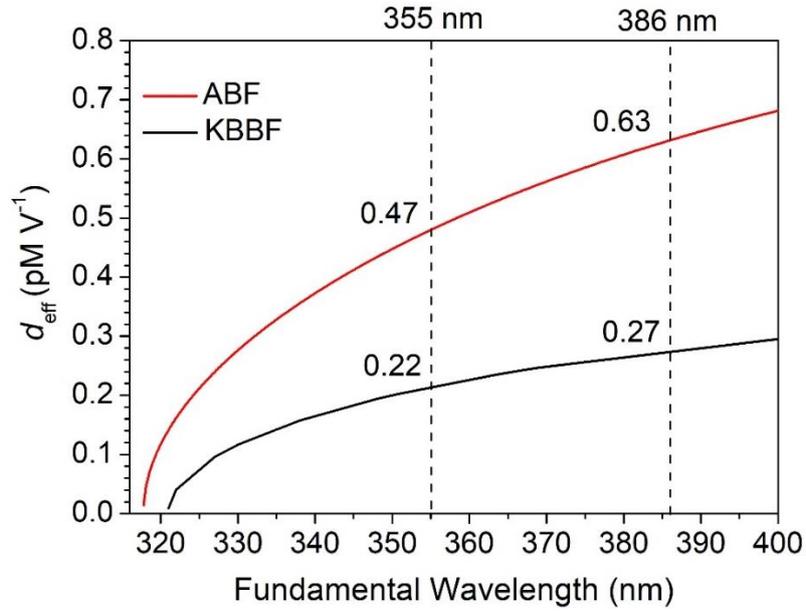

**fig. S3. Raw data of frequency doubling light for 158.9-188.0 nm using the ABF device with ($\theta$, $\varphi$) = (0 °, 70 °).**

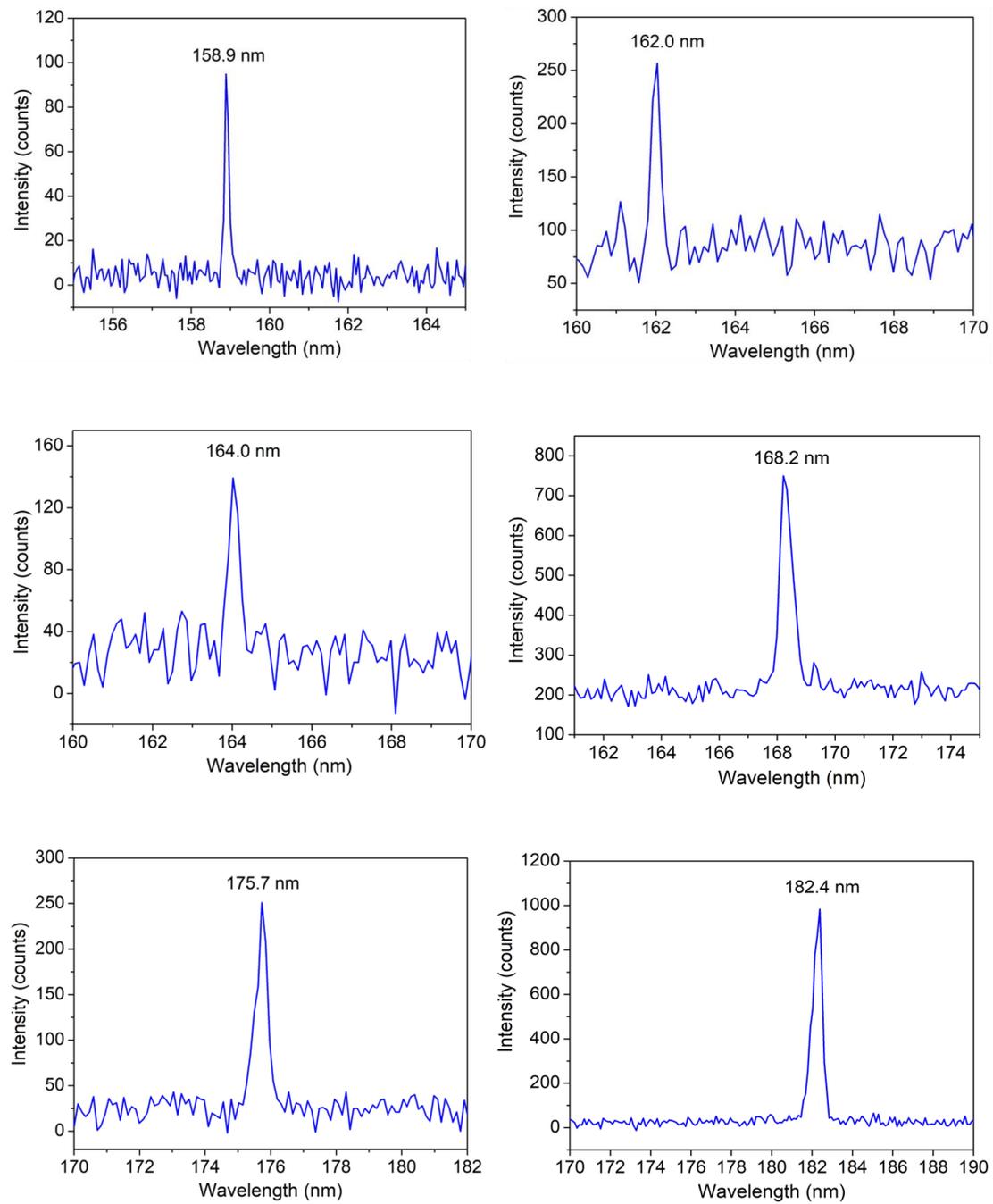

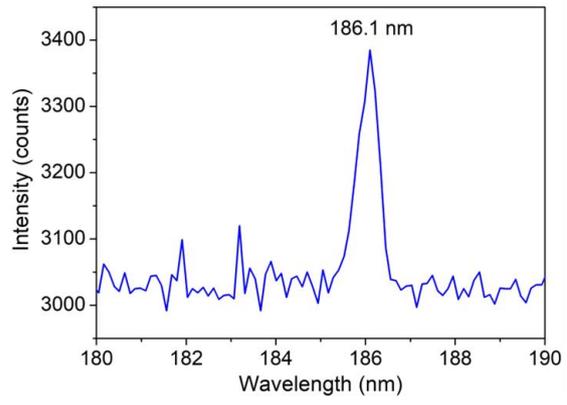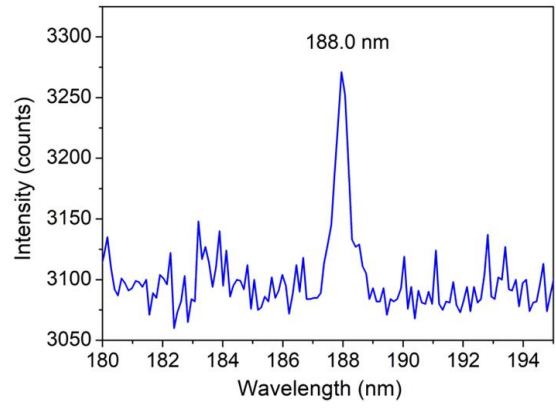

**fig. S4. Observation of frequency doubling light for 190.4-230.5 nm using the ABF device with ($\theta$, $\varphi$) = (0 °, 40 °).** The red dotted line highlights breaking through the "200nm wall". Note that the SHG intensity is normalized for a better version and the raw data are available in Fig. S5.

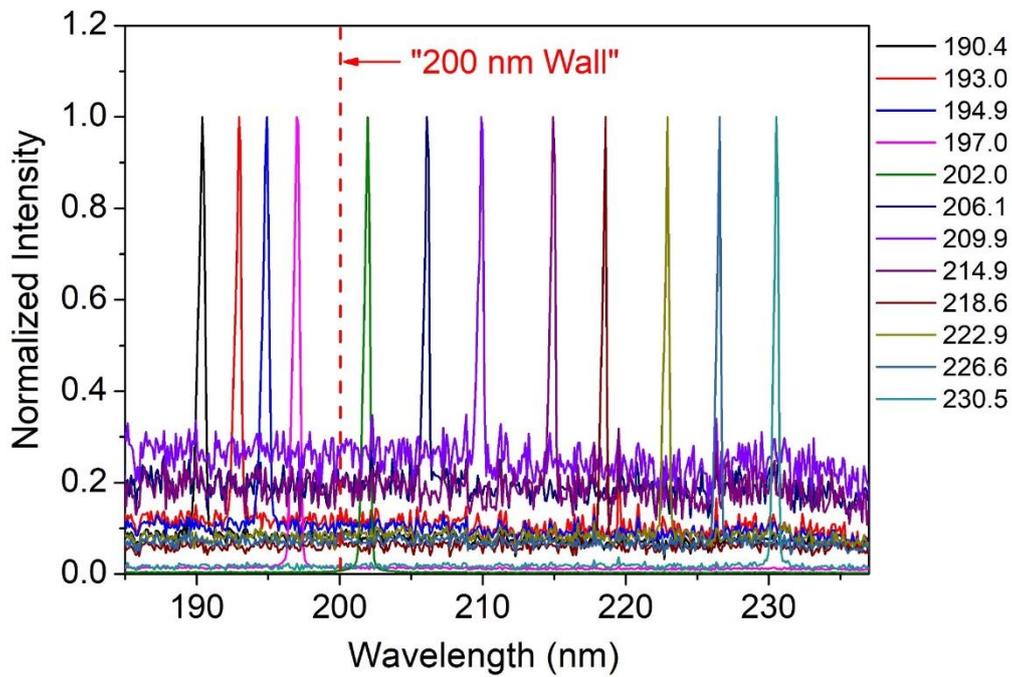

fig. S5. Raw data of frequency doubling light for 190.4-230.5 nm using the ABF device with ($\theta$, $\varphi$) = (0 °, 40 °).

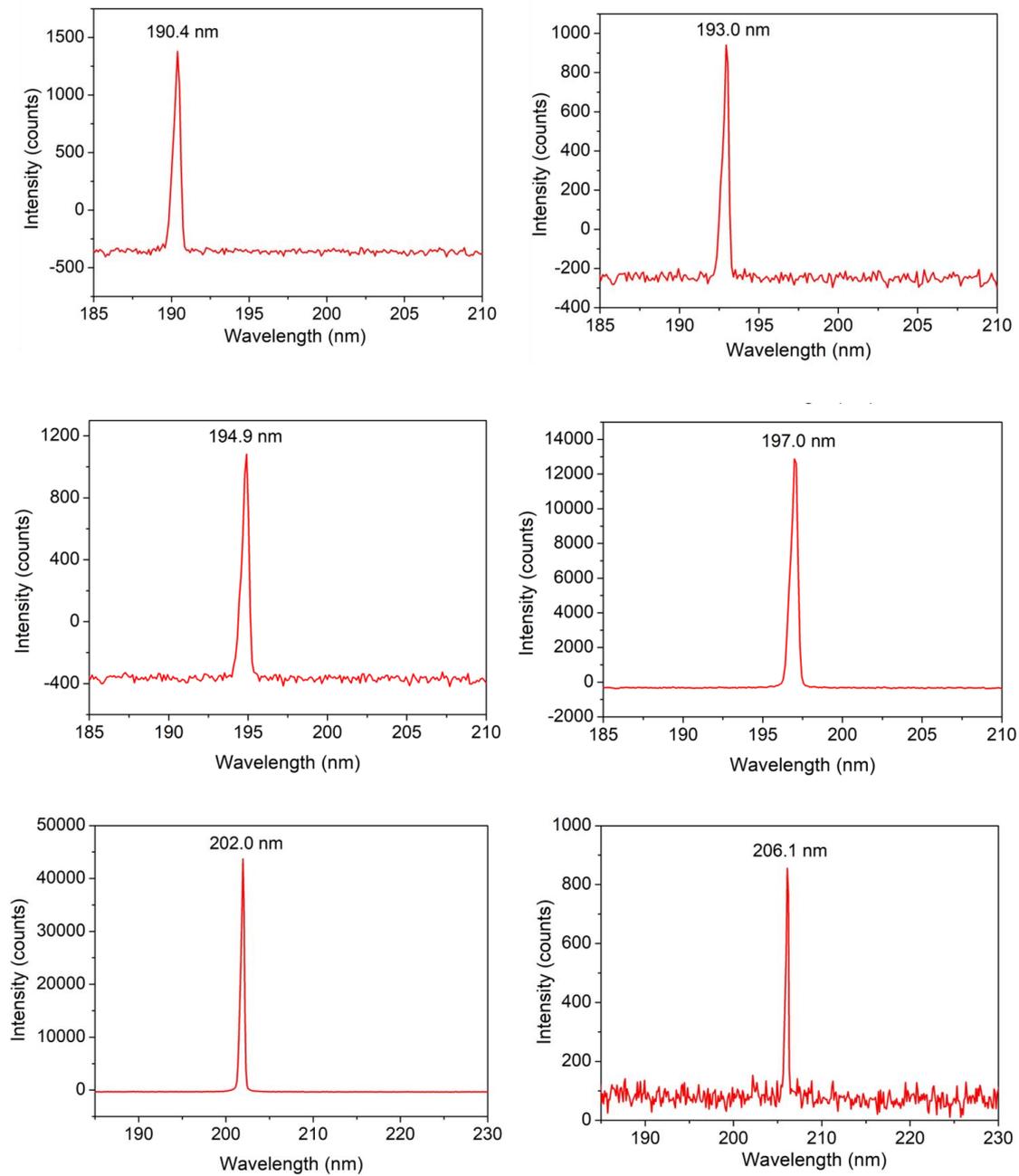

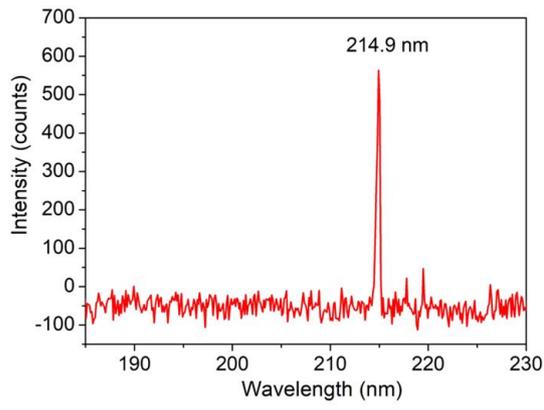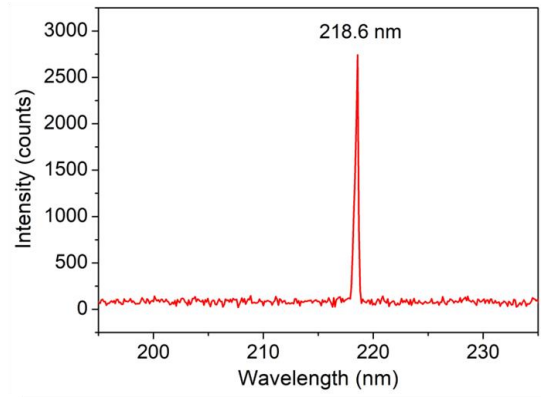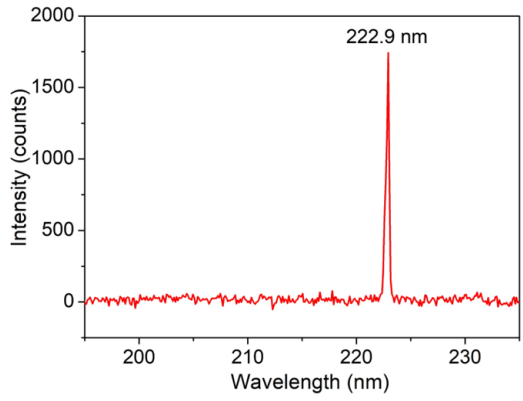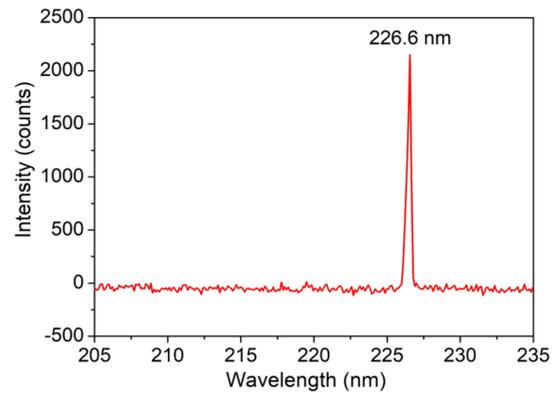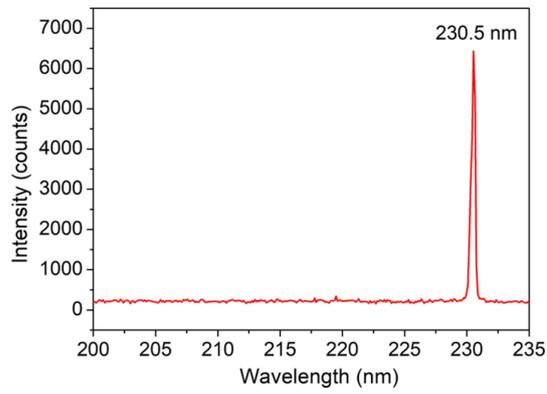

**fig. S6 Observation of frequency doubling light from 255.1-340.2 nm using the ABF device with ($\theta$, $\varphi$) = (0 °, 90 °).** Note that the SHG intensity is normalized for a better version and the raw data are available in Fig. S7. Both fundamental and SHG lights are shown.

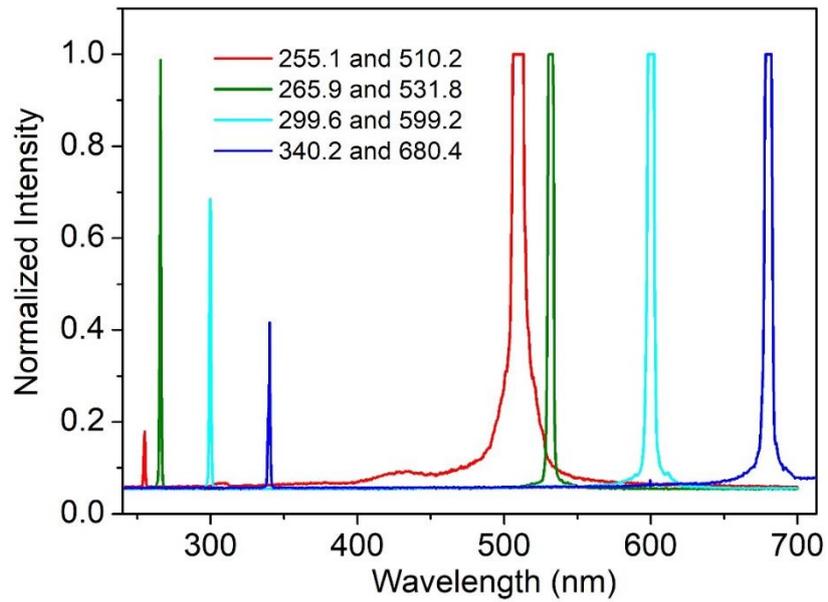

**fig. S7. Raw data of frequency doubling light for 255.1-340.2 nm using the ABF device with ($\theta$, $\varphi$) = (0 °, 90 °).** Both fundamental and SHG lights are highlighted by dot lines in the figures.

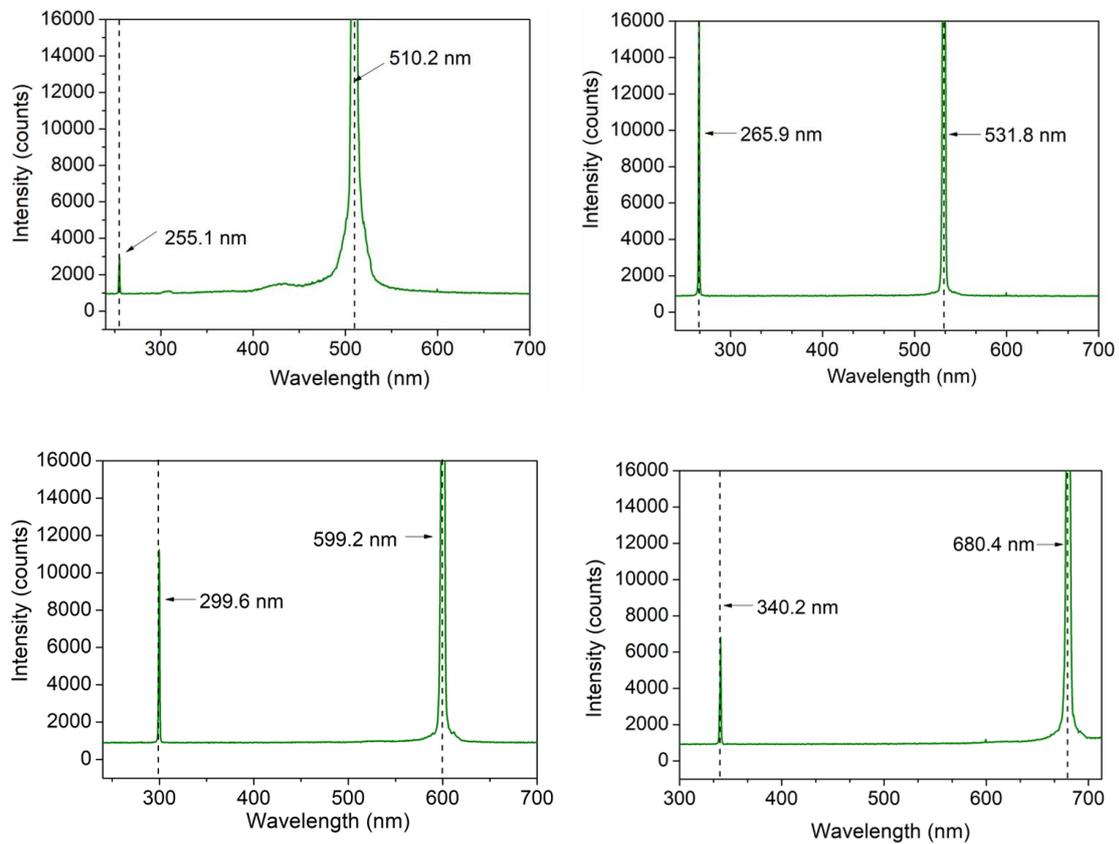

**table S1. Calculated effective SHG coefficients tensor (pm V$^{-1}$) and contribution of orbitals in the SHG coefficient $d_{32}$ of ABF.**

| Interaction orbitals | Contribution |
| --- | --- |
| NH$_4$-related orbitals (including hydrogen bond interaction orbital) | $-$ 0.229 pm V$^{-1}$ (23 %) |
| Boron and oxygen/fluorine related orbitals | $-$ 0.773 pm V$^{-1}$ (77 %) |
| Sum | $-$ 1.002 pm V$^{-1}$ |